\documentclass[5p,authoryear,preprint]{elsarticle}
\usepackage{nccmath}
\journal{Journal of High Energy Astrophysics}

\begin{document}
\begin{frontmatter}

\title{Inverse Compton emission from a cosmic-ray precursor in RX J1713.7$-$3946}

\author[]{Yutaka Ohira\corref{cor}}
\ead{ohira@phys.aoyama.ac.jp}

\author{Ryo Yamazaki}

\cortext[cor]{Corresponding author}

\address{Department of Physics and Mathematics, Aoyama Gakuin University, 5-10-1, Fuchinobe, Sagamihara 252-5258, Japan}

\begin{abstract}
Recently, the High Energy Stereoscopic System (H.E.S.S.) reported two new interesting results for a $\gamma$-ray emitting supernova remnant, RX J1713.7$-$3946 (G347.3$-$0.5). 
The first result is the establishment of a broken power-law spectrum of GeV-TeV $\gamma$-rays. 
The other is a more extended $\gamma$-ray spatial profile than the one in the X-ray band. 
In this paper, we show both of these results can be explained by inverse Compton emission from accelerated electrons. 
If the maximum energy of electrons being accelerated decreases with time, 
the broken power-law spectrum can be generated by accumulation. 
Furthermore, the extended component of $\gamma$-ray profile can be interpreted as a CR precursor of currently accelerated electrons.
\end{abstract}


\end{frontmatter}

\section{Introduction}
\label{sec:1}
Supernova remnants (SNRs) are the most plausible candidate of the origin of Galactic cosmic rays (GCRs) mainly composed of protons, electrons and nuclei. 
In fact, X-ray and $\gamma$-ray observations showed that electrons and protons (or nuclei) are accelerated in SNRs \citep{koyama95,ackermann13}. 
The diffusive shock acceleration (DSA) \citep{axford77,krymsky77,bell78,blandford78} is the most plausible acceleration mechanism of GCRs, where it is assumed that accelerated particles diffusively move around a shock. 
It predicts a power-law momentum spectrum of the accelerated particles, 
that is almost consistent with radio observations of SNRs \citep{reynolds12}. 
Another important prediction of the DSA is a CR precursor ahead of the shock front. 
Furthermore, linear analysis and several numerical simulations show that the CR precursor generates magnetic-field fluctuation \citep{bell78,bell04,niemiec08,ohira09,riquelme09,ohira10,caprioli13}.  
Recently, \citet{katsuda16} found that the upstream plasma of the SNR Cygnus loop is abruptly heated in the vicinity of shocks explained via damping of magnetic-field fluctuation in an unresolved thin CR precursor. 
However, the CR precursor of SNRs has never been directly imaged so far. 
The length scale of the CR precursor tells us the diffusion coefficient of CRs in the upstream region and includes information about magnetic-field fluctuation generated by CRs. 
The imaging observation of the CR precursor at high energies is crucial for identifying that DSA actually works at SNR shocks, thus it is eagerly anticipated.

The SNR RX J1713.7$-$3946 is one of the best studied SNRs to understand CR acceleration \cite[for a recent review, see ][]{zhang16}, detected in 
radio, X-ray, and GeV-TeV $\gamma$-ray bands. 
In particular, the origin of $\gamma$-rays from RX J1713.7$-$3946 has attracted attention over the years. 
One is the hadronic origin, that is, the $\gamma$-rays originate from accelerated protons \citep{berezhko06,yamazaki09,ellison10,zirakashvili10,inoue12,gabici14,federici15}. 
The other is the leptonic origin in which the $\gamma$-rays are generated by inverse Compton emission from accelerated electrons \citep{porter06,ellison10,zirakashvili10,finke12}. 
At present, the origin of $\gamma$-rays from RX J1713.7$-$3946 is an open problem. 
Recently, some new data about RX J1713.7$-$3946 were reported. 
\citet{katsuda15} reported the first detection of thermal X-ray line emission from RX J1713.7$-$3946 and proposed that RX J1713.7$-$3946 resulted from a type Ib/c SN. 
The High Energy Stereoscopic System (H.E.S.S.) reported two interesting results 
\citep{naurois15,abdalla16}\footnote{Details for \citet{naurois15} were provided as a full paper by \citet{abdalla16} which was submitted after this paper was submitted to arXiv.}: 
\begin{enumerate}[(1)]
\item The new GeV-TeV $\gamma$-ray spectrum breaks at about $100~{\rm GeV}$.
\item The radial profile of $\gamma$-rays is more extended than that of X-rays.  
\end{enumerate}
The former has already suggested by \citet{abdo11} although the spectrum of GeV $\gamma$-rays had a large uncertainty. 
In a simple leptonic model, the recently observed $\gamma$-ray spectrum tells us the existence of a break at a few TeV in the spectrum of accelerated electrons.  
However, it  is hardly  explained by the cooling break because it requires a strong  magnetic field or a very high photon field energy density, that conflict with other observations.  
Moreover, a one-zone leptonic model cannot explain the extended $\gamma$-ray profile. 
Therefore, a simple leptonic model seems to be confronted by the severe challenge \citep{naurois15,abdalla16}.

In this paper, we show that if the maximum energy of accelerated electrons is decreasing with time and it is now about a few TeV, the time-integrated spectrum of accelerated electrons breaks at a few TeV, that can explain the new GeV-TeV gamma ray spectrum by inverse Compton emission. 
In addition, our model can naturally explain the $\gamma$-ray profile  
by the CR precursor of currently accelerated electrons with energy of a few TeV. 
Hence, the leptonic model is still plausible and the extended component of $\gamma$-ray image is the first observation of the CR precursor of SNRs. 

\section{Time-integrated spectrum}
\label{sec:2}

\citet{ohiraetal10} showed that if the maximum energy of accelerated particles at the shock decreases with time, $E_{\max}(t)\propto t^{-\alpha}$, and the injection spectrum at the shock surface is given by ${\rm d}N/{\rm d}E{\rm d}t \propto t^{\beta-1}E^{-2}\Theta(E_{\rm max}(t)-E)$
(where $\beta>1$ and the standard test particle DSA is assumed),
then the time integrated spectrum of all accelerated particles becomes a broken power law as
\begin{eqnarray}
\frac{{\rm d}N_{\rm TI}(t,E)}{{\rm d}E}&=&\int_{0}^{t} \frac{{\rm d} N}{{\rm d}E{\rm d}t} \nonumber \\
&\propto& t^{\beta}\exp\left\{-\left(\frac{E}{E_{\rm cut}}\right)^2\right\} \nonumber \\
&\times&\left\{ \begin{array}{ll}
E^{-2}& ~(E \leq E_{\max}(t)) \\
E_{\max}(t)^{\frac{\beta}{\alpha}} E^{-\left(2+\frac{\beta}{\alpha}\right)} & ~(E \geq E_{\max}(t)) \\
\end{array} \right.,
\label{eq:bpl}
\end{eqnarray}
where no cooling is considered and $E_{\rm cut}$ is the maximum energy of accelerated particles in a whole system. 
Schematic picture of the time integrated spectrum is given by Figure~1 of \citet{ohira11}. 
Electrons above $E_{\rm max}(t)$ were accelerated in the past and make a spectrum steeper than $E^{-2}$. 
Therefore, the energy spectrum of accelerated particles inside an SNR breaks even though the radiative cooling is not significant. 
In the next Section, we provide a model that makes $\beta/\alpha=1$ and $E_{\rm max}=$ a few TeV at present, and the observed GeV-TeV $\gamma$-ray spectrum can be explained by inverse Compton emission.

It was considered in \citet{ohiraetal10} and \citet{ohira11} that particles above $E_{\max}(t)$ escape from an SNR. 
In this paper, we assume that all particles above $E_{\max}(t)$ are advected into the downstream region of SNRs. 
Particles escape from the SNR if $E_{\rm max}$ is limited by escape, while they are advected into the SNR interior if $E_{\rm max}$ is limited by a finite age of the SNR. More details are discussed in the next section.

\section{Evolution of maximum energy of accelerated electrons}
\label{sec:3} 
\begin{figure}[t]
\includegraphics[scale=0.7]{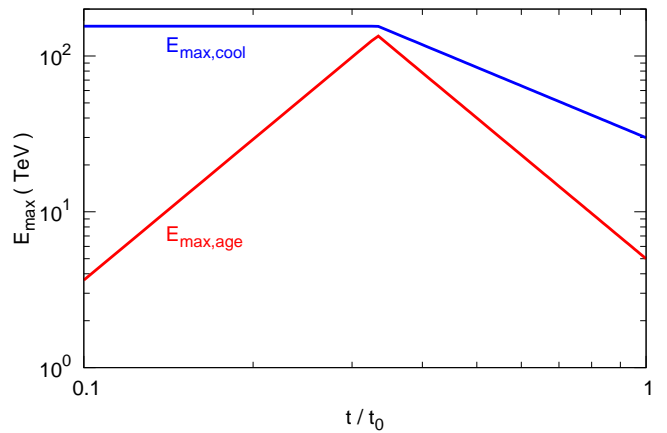}
\caption{Evolution of maximum energy of accelerated electrons $E_{\rm max}(t)$ in our model. The red and blue lines show the age and cooling limited maximum energies, respectively. $t_0$ is the SNR age at present. 
\label{fig:1}}
\end{figure}
\begin{figure*}
\begin{center}
\includegraphics[width=90mm]{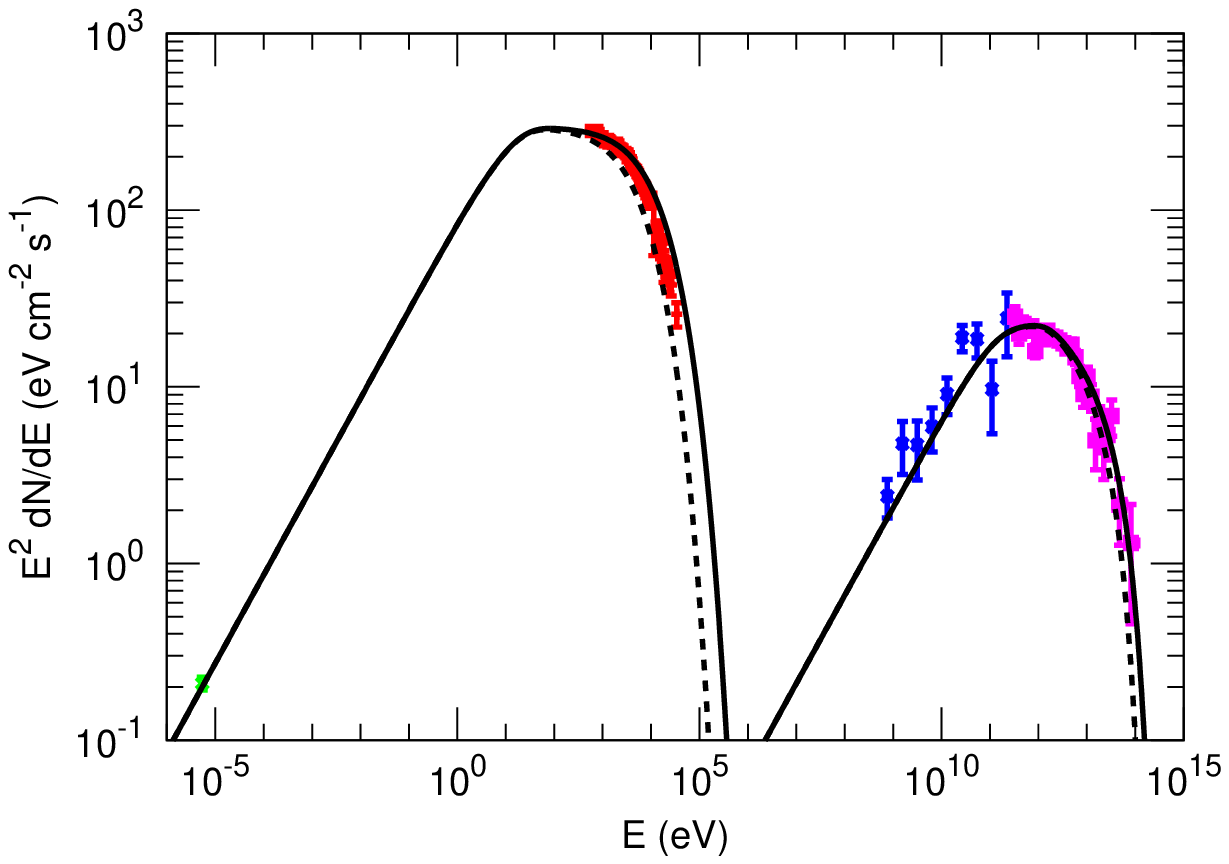}
\includegraphics[width=90mm]{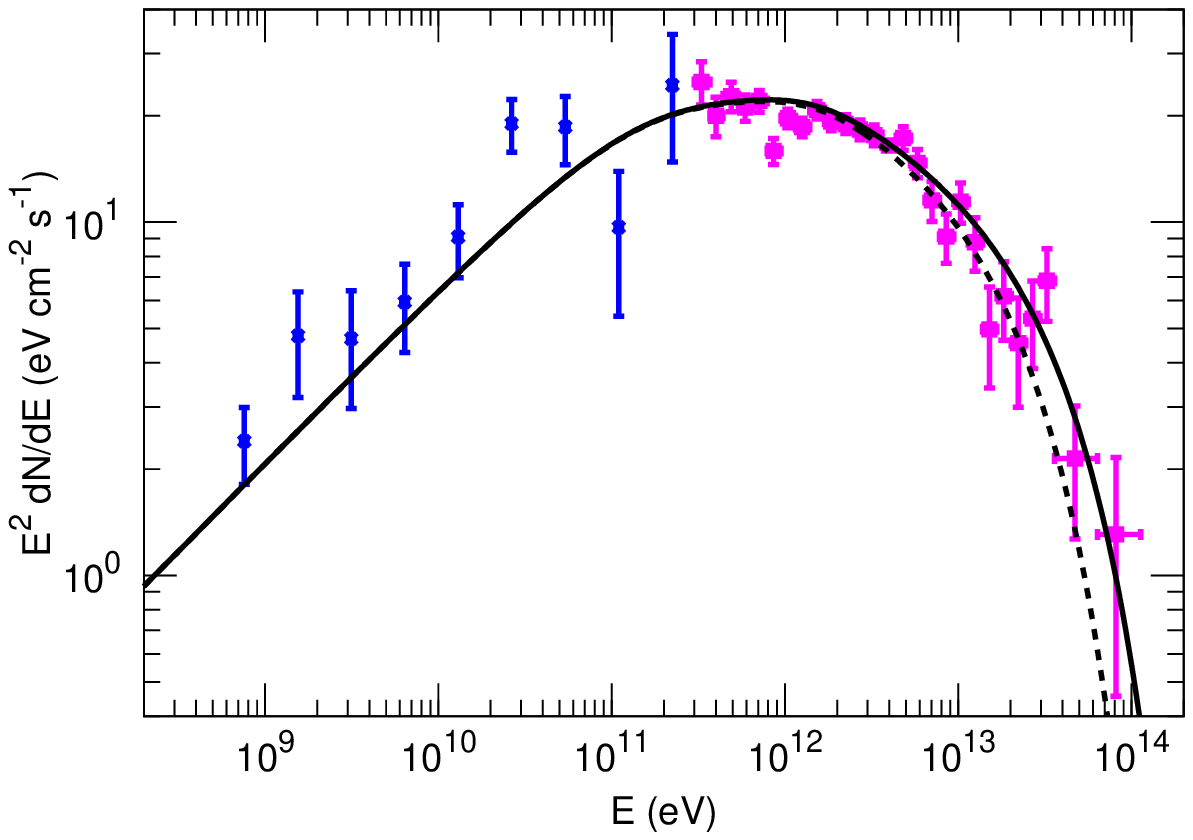}
\end{center}
\caption{Multi-band (left) and gamma-ray (right) spectra of the RX J1713.7$-$3946. 
The black solid and dashed curves show model spectra without ($E_{\rm cut} = 135~{\rm TeV}$) and with cooling ($E_{\rm cut} = 88~{\rm TeV}$) in the downstream region of SNR shocks, respectively. 
The red, green, blue, and magenta data points are given by {\it Suzaku} \citep{tanaka08}, ATCA \citep{acero09}, {\it Fermi} \citep{federici15}, and H.E.S.S. \citep{aharonian11}, respectively. 
\label{fig:2}}
\end{figure*}

Before discussing the evolution of $E_{\rm max}$ of accelerated electrons, we first set a simple model of SNR RX J1713.7$-$3946. 
The SNR is thought to be associated with the guest star AD393 \citep{wang97}. 
If so, the present age of the SNR is about $t_0 \approx1600~{\rm yr}$. 
According to the X-ray image \citep{tanaka08}, the angular radius of SNR shock is about $\theta \approx 0.45^\circ$ that corresponds to the shock radius  $r_0\approx 8~{\rm pc}$ if the distance is about $\approx 1~{\rm kpc}$ \citep{fukui03}. 
For simplicity, we here assume that the SNR is now in the free expansion phase, that is, the shock velocity is constant with time, $u_0=r_0/t_0 \approx 5000~{\rm km/s}$. 
In addition, we assume a constant density profile.  
Then, $\beta$ is 3 because the swept-up mass is proportional to $t^{3}$. 

Evolution of $E_{\rm max}$ of accelerated electrons in SNRs was discussed in \citet{ohiraetal12}. 
The $E_{\rm max}$ of electrons being accelerated is decided by a finite age, escape, or cooling. 
During the free expansion phase, the $E_{\rm max}$ does not decrease with time as long as the diffusion coefficient is spatially constant (see figures 1 and 2 of \citet{ohiraetal12} ). 
Therefore, we here assume that the diffusion coefficient around the shock front has a radial dependence, 
\begin{equation}
D =D_0 \left(\frac{E}{10~{\rm GeV}}\right)^{\delta} f(r)~~,
\end{equation}
where $f(r)$ represents radial dependence of the diffusion coefficient. 
RX J1713.7$-$3946 is thought to be expanding into a cavity produced by a stellar wind and the SNR radius is now very close to the cavity wall \citep{fukui03}. 
We expect that the outer region of the cavity has large diffusion coefficient compared with the interior, because there are many neutral hydrogen atoms and molecules in the outside region. 
A stellar wind cannot completely sweep high-density clouds into the outer region \citep[e.g. see Figure 9 of ][]{inoue12}.
Although the clouds in the stellar wind are ablated by the stellar wind plasma, some clouds would survive to the supernova explosion and some neutral particles would be present in the cavity. 
Therefore, the cavity wall (boundary between the cavity and the outer region) is actually not so sharp 
and the mean radial profile of the neutral fraction would increase with the distance from the explosion center. 
Then, the diffusion coefficient increases toward the cavity wall in the transition region. 
In order to describe such a situation, we set $f(r)$ to be 
\begin{eqnarray}
f(r) =\left\{ \begin{array}{ll}
\left(\frac{r_{\rm tr}}{r_0}\right)^{a}& ~(r \leq r_{\rm tr}) \\
\left(\frac{r}{r_0}\right)^{a} & ~(r_{\rm tr} \leq r \leq r_0) \\
1 & (r_0 \leq r)
\label{eq:d}
\end{array}\right. ~~,
\end{eqnarray}
where $r_{\rm tr}$ is a transition radius and $a>0$. 
Since the acceleration time scale is given by $t_{\rm acc}= \eta_{\rm acc}D/u_0^2$, from the condition, $t=t_{\rm acc}$, the age-limited maximum energy is given by 
\begin{eqnarray}
E_{\rm max,age} = E_{\rm max,0} \times \left\{ \begin{array}{ll}
\left( \frac{t_{\rm tr}}{t_0}\right)^{-\frac{a}{\delta}}\left(\frac{t}{t_0}\right)^{\frac{1}{\delta}} (t \leq t_{\rm tr}) \\ 
\left( \frac{t}{t_0}\right)^{\frac{1-a}{\delta}}  ~~(t_{\rm tr}\leq t \leq t_0)
\label{eq:eage}
\end{array}\right.
\end{eqnarray}
where $t_{\rm tr}=r_{\rm tr}/u_0$, and $E_{\rm max,0}=E_{\rm max}(t_0)$ is the maximum energy of currently accelerated particles and given by
\begin{equation}
E_{\rm max,0} = 10~{\rm GeV} \left(\frac{u_0^2t_0}{\eta_{\rm acc}D_0}\right)^{\frac{1}{\delta}}~~.
\end{equation}
A numerical factor $\eta_{\rm acc}$ depends on the shock compression ratio and the diffusion coefficient in the downstream region \citep{drury83}. 
Since $\delta$ is expected to be positive, 
the maximum energy decreases with time for $t_{\rm tr} \leq t \leq t_0$ if $a$ is larger than unity, and $\alpha$ becomes  
\begin{equation}
\alpha=\frac{1-a}{\delta}~~.
\end{equation}

The escape-limited maximum energy is given by the condition, $t_{\rm esc}=t_{\rm acc}$, where the escape time scale is $t_{\rm esc}= \eta_{\rm esc}r^2/D$ and $\eta_{\rm esc}$ is a numerical factor. Then, we obtain 
\begin{equation}
E_{\rm max,esc} = (\eta_{\rm acc}\eta_{\rm esc})^{\frac{1}{2\delta}}E_{\rm max,age}~~. 
\end{equation}
Since $(\eta_{\rm acc}\eta_{\rm esc})^{\frac{1}{2\delta}}$ is of the order of unity, the evolution of the escape-limited maximum energy is almost the same as the age-limited maximum energy. 
In this paper, we assume $\eta_{\rm acc}\eta_{\rm esc}>1$, so that the escape does not limit the maximum energy and all particles accelerated in the past are 
trapped in the downstream region of SNR shocks.

The synchrotron cooling time is represented by $t_{\rm cool}=9m_{\rm e}^4c^7/(4e^4B^2E)=t_{\rm c}(E/10~{\rm GeV})^{-1}$ and the condition $t_{\rm acc}=t_{\rm cool}$ gives the cooling-limited maximum energy as 
\begin{eqnarray}
E_{\rm max,cool} = E_{\rm cool,0} \times  \left\{ \begin{array}{ll}
\left(\frac{t_{\rm tr}}{t_0}\right)^{-\frac{a}{1+\delta}} & (t \leq t_{\rm tr}) \\
\left(\frac{t}{t_0}\right)^{-\frac{a}{1+\delta}} & (t_{\rm tr}\leq t \leq t_0)
\label{eq:ecool}
\end{array}\right. ~~,
\end{eqnarray}
where we assume a constant magnetic field in the downstream region because 
the shock velocity is constant in the free expansion phase. 
The cooling-limited maximum energy at the present day, $E_{\rm cool,0}$, is given by
\begin{equation}
E_{\rm cool,0} = 10~{\rm GeV} \left(\frac{u_0^2t_{\rm c}}{\eta_{\rm acc}D_0}\right)^{\frac{1}{1+\delta}}~~.
\end{equation}

Then, the evolution of the $E_{\rm max}$ of electrons being accelerated at time t, $E_{\rm max}(t)$, is given by
\begin{equation}
E_{\rm max}(t) = \min\{E_{\rm max,age},E_{\rm max,esc},E_{\rm max,cool}\}~~.
\end{equation}
In order to explain the new GeV-TeV $\gamma$-ray spectrum, we need $E_{\rm max,0}\approx$ a few TeV, and the maximum energy at $t_{\rm tr}$ is about 100 TeV.

In this paper, in oder to fit the observed spectrum, we set $a=2,~\delta=1/3,r_{\rm tr}=r_0/3$, and $\eta_{\rm acc}D_0=1.51\times 10^{27}~{\rm cm^2/s}$, so that we obtain $E_{\rm max,0}=5~{\rm TeV},~E_{\rm max}(t_{\rm tr}) = E_{\rm cut}=135~{\rm TeV}$, $\alpha=3$, and $\beta/\alpha=1$. 
These parameters are representative but not unique and other parameter sets would be allowed to explain the observed data. 
Figure~\ref{fig:1} shows the evolution of the $E_{\rm max}$ of particles being accelerated. 
The red and blue lines shows the age and cooling limited maximum energies, respectively, where we assume the temporally constant downstream  magnetic field, $B=11.5~{\rm \mu G}$ that is obtained by a spectral fitting of the X-ray and $\gamma$-ray spectra based on the leptonic model \citep{porter06,ellison10,zirakashvili10,finke12}. 
Since our model assumed that the SNR expands with a constant velocity in a uniform density medium, 
the temporally constant magnetic field strength in the downstream regions is a reasonable assumption. 
For the above parameters, the maximum energy is always limited by a finite age. 
The maximum energy increases with time up to $135~{\rm TeV}$ at $t=t_{\rm tr}$. 
Then, it decreases with time and becomes $5~{\rm TeV}$ at present. 
It should be noted that the cooling is negligible while electrons are being accelerated, 
but after advected into the downstream region, high-energy electrons lose their energy by the synchrotron cooling. 
Let $t_{\rm end}(E)$ to be the end time of the acceleration of electrons with an energy $E$. 
Then, from the condition $t_0 - t_{\rm end}(E_{\rm cut,cool}) = t_{\rm cool}(E_{\rm cut,cool})$, we can obtain a new cutoff energy by cooling, $E_{\rm cut,cool}\approx 88~{\rm TeV}$, where $B=11.5~{\rm \mu G}$ is assumed.  
For $E>E_{\rm cut,cool}$, advected electrons lose their energy in the downstream region. 
However, some electrons with an energy $E$ are in CR precursor region at $t=t_{\rm end}(E)$ and 
it takes a little time to enter the shock downstream region. 
Therefore, their residence time in the downstream region becomes smaller than $t_0-t_{\rm end}(E)$. 
In this case, we need to solve numerically the diffusion convection equation taking into account the cooling for the purpose of precisely having $E_{\rm cut,cool}$. 
In this paper, instead of solving the equation, we simply consider two cases, $E_{\rm cut}=135$ and $88$ TeV.

We calculate synchrotron and inverse Compton emissions by using the Galactic radiation field of $8~{\rm kpc}$ model of \citet{porter08}, where the Klein--Nishina effect is taken into account. 
Figure~\ref{fig:2} shows the synchrotron and inverse Compton spectra from accelerated electrons. 
The dashed and solid black curves show spectra with and without cooling in the downstream region of the SNR shock, that is, $E_{\rm cut}=135~{\rm TeV}$ and $88~{\rm TeV}$, respectively. 
Both curves are almost consistent with the observed spectrum. 
A more realistic spectrum would be between the solid and dashed curves. 
Hence, our leptonic model can explain X-ray and GeV-TeV $\gamma$-ray spectra without the cooling break. 
In this model, currently accelerated electrons cannot emit X-rays above $0.1~{\rm keV}$ and $\gamma$-rays above a few {\rm TeV} because their maximum energy is $5~{\rm TeV}$ and the downstream magnetic field strength is $11.5~{\rm \mu G}$. 
X-rays are emitted by electrons that were accelerated in the past. 
Their energy spectrum is steeper than that of currently accelerated electrons (see equation (\ref{eq:bpl})). 

\section{Radial profile of gamma rays}
\label{sec:4}
\begin{figure}
\includegraphics[scale=0.7]{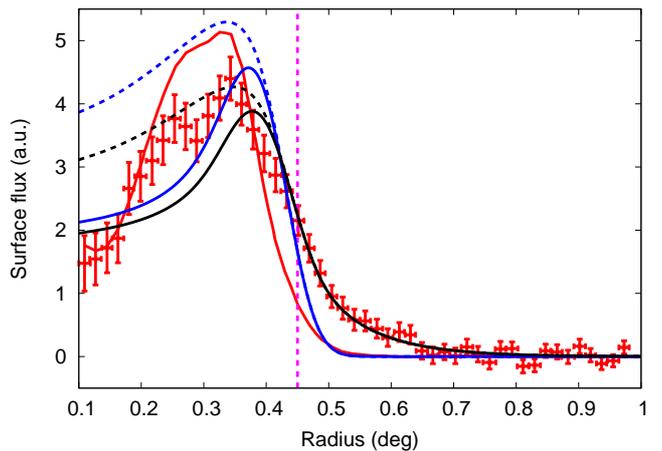}
\caption{Radial profiles of X-rays and $\gamma$-rays. 
The vertical dashed line shows the shock position expected from the X-ray image \citep{tanaka08}. 
The black and blue solid curves show model profiles of $\gamma$-rays and X-rays. 
The black and blue dashed curves show another model profiles of $\gamma$-rays and X-rays, where the step function in Equation (\ref{eq:nd}) is removed. 
The red data points and solid curve show the observed profile of $\gamma$-rays and X-rays, that are given by \citet{naurois15,abdalla16}.
\label{fig:3}}
\end{figure}

In this section, we calculate the radial profile of $\gamma$-rays. 
The DSA predicts that currently accelerated electrons make the CR precursor. 
The diffusion equation in the upstream region of an SNR shock is given by 
\begin{equation}
\frac{\partial n}{\partial t} =\frac{1}{r^2}\frac{\partial }{\partial r}\left(Dr^2\frac{\partial n}{\partial r}\right)~~.
\end{equation}
Since the SNR shock of RX J1713.7$-$3946 is now very close to the cavity wall, 
we assume that the diffusion coefficient is spatially constant in the upstream region ($r>r_0=u_0t$, see Equation~(\ref{eq:d})). 
Then, the solution for the steady-state limit ($t  \rightarrow \infty$) is given by 
\begin{equation}
n(r,t) =C\frac{r_0}{r}\exp\left\{-\frac{(r-r_0)}{D/u_0}\right\}~~, 
\label{eq:nu}
\end{equation}
where $C$ is a normalization constant. 
For a spherical shock with a constant expansion velocity, CRs distribute like the Yukawa potential in the CR precursor. 
We apply the above solution to currently accelerated particles ($E<5~{\rm TeV}$).
Highest-energy particles being accelerated at present satisfy the condition of $t_0=t_{\rm acc}$ because the maximum energy is now limited by a finite age. 
Hence, the largest length scale of the CR precursor is given by 
\begin{equation}
l_{\rm diff}(E_{\rm max}) = \frac{D}{u_0}=\frac{r_0}{\eta_{\rm acc}}~~.
\end{equation}
Interestingly, the ratio $l_{\rm diff}/r_0$ does not depend on the magnetic field but depends only on $\eta_{\rm acc}$. 

In order to calculate a downstream distribution of accelerated particles precisely, 
we need to solve numerically the diffusion convection equation with radiative and adiabatic cooling. 
Compared with the upstream region, the downstream flow is complicated because there is a contact discontinuity and recent simulations showed that the strong turbulence is generated in the downstream region by several mechanisms \citep{giacalone07,inoue09,caprioli13,ohira16a,ohira16b}. 
Then, not only the standard diffusion but also the turbulent diffusion becomes important \citep{bykov93,ohira13}. 
In this paper, we simply assume the downstream distribution ($r\leq r_0$) of currently and previously accelerated particles as follows. 
\begin{equation}
n(r,t) = C \exp\left\{-\frac{(r_0-r)}{0.25r_0}\right\} \Theta(r-0.75r_0)~~, 
\label{eq:nd}
\end{equation}
where $\Theta(x)$ is the step function, that is $\Theta(x)=1$ for $x>0$ and $\Theta(x)=0$ for $x<0$, describing a finite size of the shocked region. 
By using Equations~(\ref{eq:nu}), (\ref{eq:nd}), and the energy spectrum of accelerated electrons derived in Section~\ref{sec:3}, we calculate radial profiles of the inverse Compton and the synchrotron emissions. 
In our model, since there are no electrons with energies above $5~{\rm TeV}$ in the shock upstream region,  
X rays are not emitted from the upstream region unless the upstream magnetic field 
is larger than about $1~{\rm mG}$ which is unlikely to be realized. 
Therefore, our model does not need to calculate the X-ray profile in the upstream region. 
For the downstream X-ray profile, we assume the constant magnetic field, $B=11.5~{\rm \mu G}$.

Figure~\ref{fig:3} shows radial profiles of $\gamma$-ray and X-ray images projected on the sky. 
The black and blue solid curves are for $\gamma$-rays above $250~{\rm GeV}$ and X-rays above $1~{\rm keV}$, that are smoothed by a Gaussian function with a width of $0.05^\circ$. 
The vertical dashed line shows the shock position. 
The observed $\gamma$-ray profile extends more than the shock position and the X-ray profile, which is consistent with our model curve with $\eta_{\rm acc}=4$ (black solid curve). 
$\eta_{\rm acc}=4$ means that the downstream diffusion length scale is much smaller than that in the upstream region \citep{drury83}, 
which suggests strong magnetic field fluctuations in the downstream region. 
The shock upstream region ($\theta>0.45^\circ$) is dim in X-rays above $1~{\rm keV}$ 
because currently accelerated electrons ($E<5~{\rm TeV}$) cannot emit X-rays above $1~{\rm keV}$
but bright in $\gamma$-rays by inverse Compton emission because the currently accelerated electrons can emit $\gamma$-rays up to a few ${\rm TeV}$ and seed photons are inevitably present. 
The radial profile in the outer region ($\theta>0.45^\circ$) does not depend on 
magnetic field profiles and the downstream electron distribution. 
As an example, we show in Figure~\ref{fig:3} radial profiles of another model without the step function in Equation~(\ref{eq:nd}) (dashed curves). 
One can see that the differences appear only in the interior region.  

The observed X-ray profile is not well described by our model. 
The blue curve in Figure~\ref{fig:3} results from the assumptions of equation ~(\ref{eq:nd}) and the constant downstream magnetic field. 
In order to calculate the X-ray profile, we need to calculate the magnetic field profile and electron distribution in the downstream region, that is beyond a scope of this paper. 
\section{Discussion}
\label{sec:5}

In this paper, we assumed that all particles previously accelerated do not escape from the SNR. 
In order to supply CRs to the interstellar medium from SNRs, CRs have to escape from SNRs. In fact, $\gamma$-ray observations shows that middle-aged or old SNRs actually supply CR protons (or nuclei) to the interstellar medium  \citep{ohiraetal11,uchiyama12}. 
Hence, accelerated electrons have to escape from the SNR eventually. 
In this model, $\gamma$-rays above $250~{\rm GeV}$ from the CR precursor ($\theta>0.45^\circ$) are mainly emitted by highest-energy electrons currently  accelerated ($E=5~{\rm TeV}$). 
If higher-energy $\gamma$-rays extend more than $250~{\rm GeV}$, it could be an evidence that electrons previously accelerated ($E>5~{\rm TeV}$) have already started to escape from the SNR. 
As long as the diffusion coefficient has an energy dependence, 
the diffusion length scale has an energy dependence. 
However, it would be difficult to identify the energy dependence by current experiments because the expected energy dependence is very weak. 
The Cherenkov Telescope array \citep[CTA,][]{acharya13} will be able to observe many SNRs with better sensitivity and angular resolution, that will allow us to identify the CR precursor or escaping CR halo. 

In this paper, we consider only the acceleration of electrons. 
However, it is expected that protons and nuclei are accelerated and they produce the CR precursor. 
Therefore, the extended $\gamma$-ray profile could be explained by hadronic models \citep{zirakashvili10,federici15}. 
For the parameters adopted in this paper, the maximum energy of accelerated particles in the past does not reach $10^{15.5}~{\rm eV}$ (the knee), which is shown in Figure~\ref{fig:1}. 
However, other parameter sets, for example, a smaller $r_{\rm tr}$ makes $E_{\rm max}(t_{\rm tr})$ larger. 
If so, $\gamma$-rays originated from hadrons could appear in $100~{\rm TeV}$ range, that might be observed by CTA \citep{nakamori15}.

\section{Summary}
\label{sec:6}
In this paper, we considered the evolution of maximum energy of accelerated  electrons in the SNR RX J1713.7$-$3946 that is expanding in a cavity and their forward shock is now very close to the cavity wall. 
We assume that the diffusion coefficient around the SNR shock increases toward the cavity wall because there are neutral particles in the outside of the cavity.  
Then, the maximum energy of particles being accelerated at the shock decreases with time, 
so that an accumulated energy spectrum of accelerated electrons breaks without radiative cooling. 
We have shown that our leptonic model could explain the observed spectrum from radio to TeV $\gamma$-rays. 
In addition, our model could naturally explain the radial profile of $\gamma$-rays, which is more extended than that of X-rays, by inverse Compton emission from the CR precursor of currently accelerated electrons. 
CTA will be able to observe the CR precursor of many SNRs by direct imaging, that will open a new window on the CR physics. 

\section*{Acknowledgments}
We thank the referee for valuable comments to improve the paper. 
We also thank S. Katsuda and A. Bamba for useful comments.
This work was supported in part by Grants-in-Aid for Scientific Research of the Japanese Ministry of Education, Culture, Sports, Science and Technology No. 16K17702 (Y.O.) and 15K05088 (R.Y.).

\end{document}